\documentstyle[prl,preprint,tighten,aps,12pt,times,epsfig]{revtex}

\begin{document}
\draft

\title{ Anomalous $\gamma \to 3\pi$ amplitude in a bound-state approach }

\author{Bojan Bistrovi\' c}
\address{\footnotesize   Physics Department,
Old Dominion University, 4600 Elkhorn Avenue, 
Norfolk, VA 23529, U.S.A. , }
\address{\footnotesize  and Jefferson Lab, 12000 Jefferson Avenue, 
Newport News, VA 23606, U.S.A.}

\author{Dubravko Klabu\v{c}ar}
\address{\footnotesize Physics Department, Faculty of Science, 
   Zagreb University, Bijeni\v{c}ka c. 32, 10000 Zagreb, Croatia\\
( tel.: +385-1-4605579 ; fax: +385-1-4680336 ; e-mail: klabucar@phy.hr )}

\maketitle

\begin{abstract}
\noindent The form factor for the anomalous process 
$\gamma \pi^+ \to \pi^+ \pi^0$,
which is presently being measured at CEBAF,
is calculated in the Schwinger-Dyson approach
in conjunction with an impulse approximation.
The form factors obtained by us are compared with the ones
predicted by the simple constituent quark loop model, vector 
meson dominance and chiral perturbation theory, 
as well as the scarce already available data.

\vskip 2mm


\end{abstract}

\vspace*{6mm}

\noindent {\bf 1.} 
The Schwinger-Dyson (SD) approach to the physics of quarks and
hadrons (see Refs. \cite{N-th9807026,RW} for reviews) provides one with 
a modern constituent quark model possessing many remarkable features. 
Its presently interesting feature is its relation with the Abelian 
axial anomaly. Other bound state approaches generally have problems with
describing anomalous processes such as the $\pi^0 \to \gamma\gamma$
decay. (See Ref. \cite{KeBiKl98} for a comparative discussion thereof.)
It was therefore a significant advance in the theory of bound states,
when Roberts \cite{Roberts} and Bando {\it et al.} \cite{bando94} 
showed that the SD approach reproduces exactly (in the chiral and 
soft limit of pions of vanishing pion mass $m_\pi$) 
the famous anomalous $\pi^0\to\gamma\gamma$ ``triangle"-amplitude 
$T_{\pi}^{2\gamma}(m_\pi = 0)= e^2 N_c /(12\pi^2 f_\pi)$, and when 
Alkofer and Roberts (AR) \cite{AR96} reproduced the anomalous
``box"-amplitude for the $\gamma \to \pi^+ \pi^0 \pi^-$ process, 
in the same approach and limits. They obtained
the form factor $F_\gamma^{3\pi}(p_1,p_2,p_3)$ at the soft point,
where the momenta of all three pions 
$\{p_1,p_2,p_3\}\equiv \{p_{\pi^+},p_{\pi^0},p_{\pi^-}\}$ vanish:
\begin{equation}
F_\gamma^{3\pi}(0,0,0) \, = \, \frac{1}{e f_\pi^2} \, T_{\pi}^{2\gamma}(0) \, =
\, \frac{e N_c}{12 \pi^2 f_\pi^3} \, ,
\label{g3piAnomAmp}
\end{equation}
as predicted on fundamental grounds by Adler {\it et al.}, Terent'ev, 
and Aviv and Zee \cite{Ad+al71Te72Av+Z72}. (The number of quark colors
is $N_c=3$, while $e$ denotes the proton charge, and $f_\pi$ the pion 
decay constant.)

Just as the triangle amplitude $T_{\pi}^{2\gamma}(0)$, 
the anomalous box amplitude (\ref{g3piAnomAmp})
is in the SD approach evaluated analytically and without any fine tuning
of the bound-state description of the pions~\cite{AR96}. This happens 
because the SD approach incorporates the dynamical chiral symmetry breaking 
(D$\chi$SB) into the bound states consistently, so that the pion, 
although constructed as a quark--antiquark composite described 
by its Bethe-Salpeter (BS) bound-state vertex 
$\Gamma_{\pi}(p,k_{\pi})$, also appears 
as a Goldstone boson in the chiral limit
($k_{\pi}$ denotes the
relative momentum of the quark and antiquark constituents of the pion 
bound state). Any dependence on what precisely the solutions for 
the dynamically dressed quark propagator
\begin{equation}
S(k)= \frac{1}{i \rlap{$k$}/ \,A(k^2) + m + B(k^2)}
       \equiv  -i \rlap{$k$}/ \,\sigma_V(k^2)+\sigma_S(k^2) \, 
\label{EuclS}
\end{equation}
and the BS vertex $\Gamma_{\pi}(p,k_{\pi})$
are, drops out in the course of the analytical derivation of 
Eq.~(\ref{g3piAnomAmp}) in the chiral and soft limit. This is as 
it should be, because the amplitudes predicted by the anomaly (again
in the chiral limit $m=0=m_\pi$ and the
soft limit, {\it i.e.}, at zero four-momentum) are independent of the 
bound-state structure, so that the SD approach is the bound-state 
approach that correctly incorporates the Abelian axial anomaly.

The Abelian axial anomaly amplitudes in Eq. (\ref{g3piAnomAmp}) 
are reproduced if the electromagnetic interactions are embedded 
in the context of the 
SD approach through the framework used, for example, by 
Refs. \cite{bando94,Roberts,AR96,KeBiKl98,KeKl1,KlKe2,KeKl3},
and often called generalized impulse approximation (GIA)
- {\it e.g.}, by Refs. \cite{AR96,KeBiKl98,KlKe2,KeKl3}.
There, the quark-photon-quark ($qq\gamma$) vertex 
$\Gamma_\mu (k,k^\prime)$ is dressed so that it satisfies 
the vector Ward--Takahashi identity (WTI)
$(k^\prime-k)_\mu \Gamma^\mu(k^\prime,k)=S^{-1}(k^\prime)-S^{-1}(k)$
together with the quark propagators (\ref{EuclS}), 
which are in turn dressed consistently with the solutions for 
the pion bound state BS vertices $\Gamma_{\pi}$.
The box graph for $\gamma\to 3\pi$ in Fig. 1 is a GIA graph if
all its propagators and vertices are dressed like this.
(In the example of $\pi^0\to\gamma\gamma$, Table 1 of Ref.~\cite{KeKl1}
illustrates quantitatively the consequences of using the bare vertex
$\gamma^\mu$, which is WTI-violating in the context of the SD approach,
instead of a WTI-preserving dressed $qq\gamma$ vertex.)

In practice, one usually uses 
\cite{Roberts,AR96,KeBiKl98,KeKl1,KlKe2,KeKl3}
realistic WTI-preserving {\it Ans\"{a}tze} for $\Gamma^\mu(k^\prime,k)$.
Following AR \cite{AR96}, we employ the Euclidean form of the widely used 
Ball--Chiu ~\cite{BC} vertex, which is fully given
in terms of the quark propagator functions of Eq. (\ref{EuclS}):
 \begin{eqnarray}
        \Gamma^\mu(k^\prime,k) =
        [A(k^{\prime 2}) \! + \! A(k^2)]
       \frac{\gamma^\mu}{\textstyle 2}
        + \frac{\textstyle (k^\prime+k)^\mu }
               {\textstyle (k^{\prime 2} - k^2) }
        \{[A(k^{\prime 2}) \! - \! A(k^2)] \,
        \frac{\textstyle ({\rlap{$k$}/}^\prime + \rlap{$k$}/) }{\textstyle 2}
         - i [B(k^{\prime 2}) \! - \! B(k^2)] \, \}~.
        \label{BC-vertex}
        \end{eqnarray}

The amplitude $T_{\pi}^{2\gamma}$ obtained in the chiral and soft limit 
is an excellent approximation for the realistic $\pi^0\to\gamma\gamma$
decay. On the other hand, the already published \cite{Antipov+al87} 
and presently planned Primakoff experiments at CERN \cite{Moinester+al99},
as well as the current CEBAF measurement of the
$\gamma \pi^+ \to \pi^+ \pi^0$ process \cite{Miskimen+al94}
involve values of energy and momentum transfer sufficiently
large to give a lot of motivation for theoretical predictions 
of the extension of the anomalous $\gamma\to 3 \pi$ amplitude 
away from the soft point. In the present paper we follow 
essentially the approach of AR \cite{AR96}, the 
difference being precisely the way in which the $\gamma\to 3 \pi$ 
form factor is extended beyond the soft point. We perform this 
extension guided by the insights from our Ref.~\cite{BiKl99PRD}.

\vspace*{3mm}

\noindent {\bf 2.}
Considering just one graph, for example Fig. 1, enabled 
Ref.~\cite{AR96} to reproduce analytically the anomalous 
amplitude~(\ref{g3piAnomAmp}) for $p_1=p_2=p_3=0$. However, 
computing the form factor $F_\gamma^{3\pi}$ beyond the soft 
limit requires 
careful inclusion of all six contributing graphs, obtained 
from Fig. 1 by the permutations of the vertices of the three different 
pions $\pi^a=\pi^+, \pi^0, \pi^-$.
Otherwise, $F_\gamma^{3\pi}$ would not be properly symmetrical under 
$p_1 \leftrightarrow p_2 \leftrightarrow p_3$.
In Fig. 1, as well as in the other five associated graphs, the relative 
momenta of the constituents of the pion bound states,
as well as the momenta flowing through the four sections of the quark loop,  
are conveniently given by various combinations of the symbols
$\alpha, \beta, \gamma = +, 0, -$ in $k_{\alpha\beta\gamma} \equiv 
k + (\alpha p_{1} + \beta p_{2} + \gamma p_{3})/2$.

If we denote the contribution of the first diagram (Fig. 1) by
$-i\varepsilon^\mu\:\epsilon_{\mu\nu\rho\sigma}\: p_1^\nu p_2^\rho p_3^\sigma
 f^{3\pi}_\gamma(p_1,p_2,p_3)$, 
where $\varepsilon^\mu$ is the photon polarization vector,
the $\gamma\to 3\pi$ amplitude ${\mathcal{A}}_{\gamma}^{3\pi}$,
{\it viz.}, the total scalar form factor  
$F_\gamma^{3\pi}(p_{1},p_{2},p_{3})$ associated with it, 
is written as 
\begin{eqnarray}
{\mathcal{A}}_{\gamma}^{3\pi} &=& - i\varepsilon^\mu\:
\epsilon_{\mu\nu\rho\sigma}\: 
p_{1}^\nu p_{2}^\rho p_{3}^\sigma \: 
F_\gamma^{3\pi}(p_{1},p_{2},p_{3})  \nonumber\\
&=&
- i \varepsilon^\mu\:  \epsilon_{\mu\nu\rho\sigma}\: p_{1}^\nu
p_{2}^\rho p_{3}^\sigma \:
f_\gamma^{3\pi}(p_{1},p_{2},p_{3}) 
+\left[\mbox{all permutations of }\pi^+(p_1), \pi^0(p_2), \pi^-(p_3)\right] \, .
\label{g3piVertex}
\end{eqnarray}

In Ref. \cite{BiKl99PRD} we computed (\ref{g3piVertex}), {\it i.e.}, the
form factor $F_\gamma^{3\pi}$, in the ``free" quark loop (QL) model 
(and hence also the lowest order $\sigma$-model and chiral quark models) 
with the constant constituent mass $M$. In the SD approach, one instead 
has the momentum-dependent (Euclidean) quark mass function 
$M(k^2)\equiv B(k^2)/A(k^2)$. The functions $A(k^2)$ and $B(k^2)$, 
{\it i.e.}, the dressed quark propagators (\ref{EuclS}), are in 
principle the solutions of the appropriate SD equation. The 
quark-pion vertices $\Gamma_{\pi}(p,k_{\pi})$
are the bound-state vertices obtained as the pion solutions of the
BS equation consistently coupled with the SD equation for the quark
propagator through the usage of its solution $S(k)$ and the same 
interaction. (See Refs. \cite{jain93b,MarisRoberts97PRC56} and 
references therein for 
examples thereof, and Refs. \cite{N-th9807026,RW,KeBiKl98} for reviews 
and applications.) This approach is therefore also often called the 
coupled SD-BS approach ({\it e.g.}, by Refs. \cite{KeBiKl98,KlKe2,KeKl3}).

However, in the variant of the SD approach used by Roberts and 
Alkofer~\cite{Roberts,AR96}, they avoided solving the SD equation 
for the dressed quark propagator $S$ by using a phenomenologically
realistic {\it Ansatz} for the dressed quark propagator (\ref{EuclS}). 
In principle, one could invert such a propagator Ansatz and find 
out which interaction would give rise to it through the SD equation.
Then, owing to working in the chiral {\it and} soft limit
they also automatically obtained the solution of the BS equation.
In this limit, when the chiral symmetry is not broken explicitly 
by $m\neq 0$, but only dynamically, and when pions must consequently 
appear as Goldstone bosons, the solution for the pion
bound-state vertex $\Gamma_{\pi}$, corresponding to the Goldstone pion,
is -- to the order ${\cal O}(p^0)$ -- given 
by the dressed quark propagator $S(k)$ (\ref{EuclS}).
For the pion bound-state vertex 
$\Gamma_{\pi}$, Ref. \cite{Roberts,AR96}
concretely used the solution, given in Eq. (\ref{ChLimSol}) immediately
below, that is of zeroth order in the pion momentum $p$.
This is appropriate close to the soft limit $p^\mu \to 0$.
The chiral and soft limit $\Gamma_{\pi}$, Eq. (\ref{ChLimSol}),
fully saturates the Adler-Bell-Jackiw axial anomaly \cite{bando94,Roberts}.
In the chiral limit, the pion decay constant $f_\pi$ is found \cite{JJ}
to be equal to the normalization constant of $\Gamma_{\pi}$,
whereas its ${\cal{O}}(p^0)$
piece is proportional to the chiral-limit solution for $B(k^2)$ from 
Eq.~(\ref{EuclS}):
\begin{equation}
\Gamma_{\pi}(p^2\! =\! - m_\pi^2\! =\! 0; k)\!
\equiv \Gamma_{\pi}(k)\!  =\! 
\frac{ B_{0}(k^2)}{f_\pi} \, \gamma_5 \, .
\label{ChLimSol}
\end{equation}
The propagator function $B_{0}(k^2) \equiv B(k^2)_{m=0}$ is the one
obtained in the chiral limit of the vanishing current quark mass $m$,
where the quark constituent mass arises purely from D$\chi$SB. 
Eq. (\ref{ChLimSol}) is analogous to the quark-level Goldberger-Treiman
relation $g = M/f_{\pi}$ for ``free" constituent quarks with the 
constant mass $M$. In that simpler case, the constant quark-pion 
pseudoscalar point coupling $g \gamma_5$ corresponds to the pion 
BS vertex $\Gamma_{\pi}(p,k_{\pi})$ in the SD approach. 

Since the pion is in a good approximation an (almost) massless
Goldstone boson, we follow AR~\cite{Roberts,AR96} 
in approximating the BS-vertex of the 
realistically massive pion by Eq. (\ref{ChLimSol}): 
\begin{equation}
\Gamma_{\pi}(p^2\! =\!- m_\pi^2\! =
\! -[m^{\scriptscriptstyle exp}_\pi]^2; k)\! \approx \Gamma_{\pi}(k) \, .
\end{equation}

The contribution of the single diagram Fig. 1, 
$ - i\epsilon_{\mu\nu\rho\sigma}\: p_1^\nu p_2^\rho p_3^\sigma
 f^{3\pi}_\gamma(p_1,p_2,p_3)$, is then 
\begin{eqnarray}\label{eq1}
- \int\frac{\,d^4 k}{(2\pi)^4}\, {\mathrm{Tr}} \left\{
\, i \, e \, {Q} \,
\Gamma_\mu(k_{\scriptscriptstyle +++},k_{\scriptscriptstyle ---}) \;
S(k_{\scriptscriptstyle ---})\;\sqrt{2}\, \tau_+ \,
\Gamma_{\pi}(k_{\scriptscriptstyle --0})
\,\,  \qquad \qquad \qquad
\right.\nonumber \\  \left. \qquad  \qquad  \quad \times
S(k_{\scriptscriptstyle --+}) \;\tau_3 \, 
\Gamma_{\pi}(k_{\scriptscriptstyle -0+})
\; S(k_{\scriptscriptstyle -++}) \; \sqrt{2}\, \tau_- \,
\Gamma_{\pi}(k_{\scriptscriptstyle 0++}) \; S(k_{\scriptscriptstyle +++})
\; \right\} \, ,
\end{eqnarray}
where the Pauli $SU(2)$ matrices $\tau_3$ and 
$\tau_\pm = (\tau_1 \pm i\tau_2)/2$ correspond, respectively, 
to $\pi^0$ and emitted $\pi^\mp$ (or absorbed $\pi^\pm$).
The quark charge matrix in the $SU(2)$-isospin space is 
${Q}\equiv\mbox{\rm diag}[Q_u,Q_d] = \mbox{\rm diag} [2/3,-1/3]$.
For this particular diagram 
the isospin trace is 
${\mathrm{Tr}}\left({Q} \tau_+ \tau_3 \tau_-\right) = 
(-1) Q_u = -2/3$.
The color trace yields the factor $N_c$.  The Dirac trace leads to the form 
\begin{eqnarray}
{\mathrm{Tr}} \{\ldots\} &=& T_1 \epsilon_{\mu\nu\rho\sigma}\: p_1^\nu p_2^\rho
p_3^\sigma
      + T_5 k^\mu\epsilon_{\alpha\nu\rho\sigma}\: k^\alpha p_1^\nu
      p_2^\rho p_3^\sigma \nonumber \\
& &   + T_2 \epsilon_{\mu\nu\rho\sigma}\: k^\nu p_2^\rho p_3^\sigma
      + T_3 \epsilon_{\mu\nu\rho\sigma}\: p_1^\nu k^\rho p_3^\sigma
      + T_4 \epsilon_{\mu\nu\rho\sigma}\: p_1^\nu p_2^\rho k^\sigma\; ,
\end{eqnarray}
where $T_n$'s are functions of scalar products $p_i\cdot p_j$
and $k\cdot p_j$ only ($i,j = 1,2,3$). They are lengthy expressions
to be integrated over the loop momentum $k$, so we do not
present them here explicitly. Obviously, evaluating $F_\gamma^{3\pi}$ 
in the SD approach, is a harder task than in the context of our
earlier Ref. \cite{BiKl99PRD} where the quark-pion coupling is
constant instead of the present BS-vertex (\ref{ChLimSol}), and 
the quark propagator has a constant constituent mass, as opposed 
to Eq.~(\ref{EuclS}). Nevertheless, it is possible to formulate an 
expansion in the pion momenta similar to that in Ref. \cite{BiKl99PRD}.
The $T_n$-functions ($n = 1, ...,5$) are expanded
around the soft limit $p_i=0$,
\begin{equation}
f(k,p_i)= f(k,0) + \sum_i p^\mu_i
\left.\left[\frac{\partial f(k,p_i)}{\partial p^\mu_i}\right]\right|_{p^\mu_i=0} +
\frac{1}{2}\sum_{i,j}p^\mu_i\,p^\nu_j
\left.\left[\frac{\partial^2 f(k,p_i)}{\partial p^\mu_i \,\partial
p^\nu_j}\right]\right|_{p^\mu_i=0} +\ldots   \, ,
\end{equation}
whereby the problem is reduced to evaluating integrals over the loop 
momentum $k$ which contain in their integrands only functions of $k^2$ 
times powers of scalar products $k\cdot p_j$. The integrals with an odd 
number of $k_\mu$ factors vanish, while the integrals with an even number 
of $k_\mu$'s are turned into integrals over pure functions of $k^2$ 
through symmetric integration, {\it i.e.}, by utilizing 
\begin{eqnarray}
\int k^\mu k^\nu f(k^2) \,d^4 k
      &=&\frac{g^{\mu\nu}}{4} \int k^2 f(k^2) \,d^4 k \\
\label{eq9}
\int k^\mu k^\nu k^\alpha k^\beta f(k^2)\,d^4 k
      &=& \frac{g^{\mu\nu}g^{\alpha\beta}
      + g^{\mu\alpha}g^{\nu\beta} + g^{\mu\beta}g^{\alpha\nu}}{24}
      \int (k^2)^2 f(k^2) \,d^4 k\; .
\end{eqnarray}
Conveniently defining 
\begin{equation}
f^{3\pi}_\gamma(p_1,p_2,p_3) \equiv- \frac{e\, N_c}{2\pi^2 \,f_\pi^3}
{\mathrm{Tr}}\left({Q} \tau_+ \tau_3 \tau_-\right) J(p_1,p_2,p_3) \, ,
\end{equation}
and analogously for the other diagrams,
the $\gamma 3\pi$ form factor
written as the sum over the six diagrams is
\begin{eqnarray}
F_\gamma^{3\pi}(p_1,p_2,p_3)&=& \frac{e\, N_c}{2\pi^2 \,f_\pi^3}
\left(\frac{2}{3} \left\{
J(p_1,p_2,p_3) + J(p_1,p_3,p_2) + J(p_2,p_1,p_3) \right\}
\right. \nonumber \\ && \left.
 -\frac{1}{3} \left\{
J(p_3,p_1,p_2) + J(p_3,p_2,p_1) + J(p_2,p_3,p_1) \right\}  \right) \, .
\label{sumAmp}
\end{eqnarray}

Owing to our expansion method, $J(p_1,p_2,p_3)$ 
and its companions with the permuted arguments $p_1,p_2,p_3$,
are given in terms of expansions in the scalar products of
the external momenta $p_1,p_2,p_3$, and the coefficients are 
given by integrals of functions (coming from the propagators and vertices)
of the squared loop momenta $k^2=\ell$. We evaluate these integrals
in Euclidean space. For example, consider the lowest, zeroth 
order contribution to the expansion, $J(0,0,0)$, which determines
the $\gamma 3\pi$ amplitude at the soft point.
It is given by the loop integral
\begin{eqnarray}
J(0,0,0) &=&  \int_0^\infty \,d \ell\: \ell\: {B_0(\ell)}^3 \,
\sigma_V(\ell)^2 \left[  \,A(\ell)\,\sigma_S(\ell)\,\sigma_V(\ell)
\right.\nonumber \\
& & \left.  +\, \frac{1}{2} \,\ell\,\sigma_S(\ell)\,\sigma_V(\ell)\,A'(\ell)
  - \, \frac{1}{2} \ell{\,\sigma_V(\ell)}^2\,B'(\ell) -
\, \frac{3}{2}\, \ell \,A(\ell)\,\sigma_V(\ell)\,\sigma_S'(\ell) \right]  \, .
\label{J000}
\end{eqnarray}

The two equivalent pairs of functions in the Euclidean quark 
propagator~(\ref{EuclS}) are connected with each other through 
the relations 
\begin{equation}
A(k^2)=\frac{\sigma_V(k^2)}{k^2 \, \sigma_V^2(k^2) + \sigma_S^2(k^2)}
 \, \,  , \qquad
B(k^2)=\frac{\sigma_S(k^2)}{k^2 \, \sigma_V^2(k^2) + \sigma_S^2(k^2)} - m
\, .
\end{equation}
In our present convention, $m$ is separated out of $B(k^2)$
which is thus purely dynamically generated
in contrast to the convention we
used previously \cite{KeBiKl98,KeKl1,KlKe2,KeKl3} where the quark mass 
$m$ which breaks chiral symmetry explicitly was lumped into $B(k^2)$.

In the chiral limit, where not only $B_0(\ell)^3$
but {\it all} propagator functions (\ref{EuclS}) appearing
in Eq. (\ref{J000}) correspond to the $m=0$ case, 
AR \cite{AR96} evaluated $J(0,0,0)$ analytically: 
its value {\it in the chiral limit}, $J_0 \equiv J_0(0,0,0)$, 
is always $J_0 = 1/6$ irrespectively of what the functions defining the
quark propagator (\ref{EuclS}) and the pion BS vertex (\ref{ChLimSol})
concretely are. This enabled AR to prove that, remarkably, the 
SD approach exactly
reproduces the soft-point amplitude (\ref{g3piAnomAmp}) independently 
of details of bound state structure. Thus, this bound-state approach  
consistently incorporates not only the ``triangle", but also the 
``box" axial anomaly. 

Since $J(0,0,0)$ is equal in every diagram, and in the chiral limit it is 
always $J_0 = 1/6$, our sum over diagrams (\ref{sumAmp}) also reproduces 
the chiral-limit result (\ref{g3piAnomAmp}) for $F_\gamma^{3\pi}(0,0,0)$.

Same as in Ref. \cite{BiKl99PRD}, we found having the sum of the diagrams
essential for obtaining the correct $\gamma 3\pi$ amplitude beyond the 
soft point, where different diagrams contribute different combinations 
of powers of the scalar products $p_i \cdot p_j$.
To get $F_\gamma^{3\pi}(p_1,p_2,p_3)$ symmetric under the interchange
of the three external momenta, one needs to consider the sum of at
least three graphs corresponding to one of the combinations enclosed
in the curly brackets in Eq. (\ref{sumAmp}). As in the simpler case
of the ``free" constituent quark loop calculation, these curly brackets
are equal to each other. 

\vspace*{3mm}

\noindent {\bf 3.}
Unlike $J_0$, the expansion coefficients of terms beyond soft and 
chiral limits are not independent on the internal structure of the pion.
To evaluate the integrals giving 
them, we must specify the propagator
functions in Eqs. (\ref{EuclS}) and (\ref{ChLimSol}). 
We adopt the AR quark propagator {\it Ans\" atze} 
supposedly suitable for modeling confined quarks \cite{AR96}, namely 
\begin{eqnarray}
\bar\sigma_S(x) &=& C_{\bar{m}} \,e^{-2x}+2\bar{m}
	\frac{1-\displaystyle{e^{-2(x+\bar{m}^2)}}}{2(x+\bar{m}^2)}+
	\frac{1-\displaystyle{e^{-b_1 x}}}{b_1 x}
	\frac{1-\displaystyle{e^{-b_3 x}}}{b_3 x}\left(
	b_0+b_2\frac{1-\displaystyle{e^{-\epsilon x}}}{\epsilon x}\right) \, \,  , \\
\bar\sigma_V(x)&=& \frac{2(x+\bar{m}^2)-1+e^{-2(x+\bar{m}^2)}}
{2(x+\bar{m}^2)^2} - \bar{m}\:C_{\bar{m}}\,e^{-2x}   \,\,  ,
\end{eqnarray}
where the dimensionless functions $\bar\sigma_S(x)$ and $\bar\sigma_V(x)$
are related to the scalar and vector propagator functions through the 
characteristic mass scale $\Lambda = \sqrt{2D}$:
\begin{equation}
\bar\sigma_S(x) = \sqrt{2D}\: \sigma_S(k^2)  \,  , 
	\qquad
\bar\sigma_V(x) = 2D\: \sigma_V(k^2)    \,  ,   
\label{connsigma}
\end{equation}
along with the quark momentum and mass,
$k^2 = \ell = 2D\:x$ and $\bar{m}={m}/{\sqrt{2D}}$.

The above quark propagator  {\it Ans\" atze}, together with the 
chiral-limit pseudoscalar BS vertex (\ref{ChLimSol}),
define the model of the quark substructure of the
light pseudoscalar meson -- the pion. By fitting a considerable
number of pion observables, 
the parameters were fixed \cite{AR96} to the values  
\begin{equation}
\begin{array}{rclcrclcl}
C_{\bar{m}} &=& 0.121 & \qquad & \bar{m} &=& 0 & \leftarrow & \mbox{in the 
chiral limit} \\
C_{\bar{m}} &=& 0 & \qquad & \bar{m} &=& 0.00897 & \leftarrow & \mbox{for
massive quarks}\\
b_0 &=& 0.131 & & b_1 &=& 2.90 & \qquad &\epsilon = 10^{-4}  \\
b_2 &=& 0.603 & & b_3 &=& 0.185 & \qquad &D = 0.16\, \mbox{GeV}^2  \, .  
\end{array}
\end{equation}
We present our expansion around the soft point by re-writing the
(dimensionful) expansion coefficients in terms of dimensionless
numbers divided by the appropriate power of a characteristic mass 
scale $\Lambda$. In the free constituent quark loop calculation 
\cite{BiKl99PRD}, this scale was of the order of the constituent 
quark mass $M$. In the present model, it is obviously 
$\Lambda =  \sqrt{2 D} = 565.69 \, \mathrm{MeV}$. 
After introducing the form factor normalized
to the anomaly amplitude (\ref{g3piAnomAmp}),
${\widetilde{F}}^{3\pi}_\gamma(p_1,p_2,p_3) \equiv
F^{3\pi}_\gamma(p_1,p_2,p_3)/F^{3\pi}_\gamma(0,0,0)$,
our expansion \emph{for general, possibly off-shell impulses} $p_i$, 
to the order ${\cal O}(p^4)$ becomes
\begin{eqnarray}\label{amplRA1}
\widetilde{F}^{3\pi}_\gamma(p_1,p_2,p_3) &=& 0.96274 - 
\frac{0.94952}{\Lambda^2 } (p_1\cdot p_2+  p_1\cdot p_3+ p_2\cdot p_3)
        - \frac{0.84175}{\Lambda^2}  (p_1^2 + p_2^2+ p_3^2 )
\nonumber \\ &+&
\frac{0.63155}{\Lambda^4} \left( (p_1\cdot p_2)^2+(p_1\cdot
p_3)^2+(p_2\cdot p_
3)^2 \right)
\nonumber \\ &+&
\frac{0.76839}{\Lambda^4} \left(  p_1^2 \, p_2^2 +p_1^2 \, p_3^2 
+ p_2^2 \, p_3^2\right)
+\frac{0.44218}{\Lambda^4}\left(  p_1^4+p_2^4+p_3^4\right)
\nonumber \\ &+&
\frac{1.02189}{\Lambda^4} \left(  p_1\cdot p_2 \, p_1\cdot p_3+
 p_1\cdot p_2 \, p_2\cdot p_3+p_1\cdot p_3 \, p_2\cdot p_3 \right)
\nonumber \\ &+&
\frac{0.97567}{\Lambda^4}\left(  p_1^2 \, p_1\cdot p_2  
+p_1^2 \, p_1\cdot p_3+
     p_2^2 \, p_1\cdot p_2  +p_2^2 \, p_2\cdot p_3+
     p_3^2 \, p_1\cdot p_3  +p_3^2 \, p_2\cdot p_3\right)
\nonumber \\ &+&
\frac{0.83507}{\Lambda^4}\left( p_1^2 \, p_2\cdot p_3 +
p_2^2 \, p_1\cdot p_3 +p_3^2 \, p_1\cdot p_2 \right) + {\cal O}(p^6) \, .
\end{eqnarray}
\par

Since this was obtained with the propagators in the presence of a 
small ($\bar{m}=0.00897$) explicit chiral symmetry breaking, the 
zeroth-order term [$ 6 J(0,0,0) = 0.96274$] slightly differs from 1. 
Note the difference with respect to the simpler ``free"
quark loop case \cite{BiKl99PRD}, where the zeroth-order term 
$6 M^4 I(0,0,0) = 1$ always. For the chiral quark propagators,
\begin{eqnarray}\label{amplRAK1}
\widetilde{F}^{3\pi}_\gamma(p_1,p_2,p_3)_0 &=& 1 - 
\frac{0.92228}{\Lambda^2 } (p_1\cdot p_2+  p_1\cdot p_3+ p_2\cdot p_3)
      - \frac{0.83476}{\Lambda^2}  (p_1^2 + p_2^2+ p_3^2 )
\nonumber \\ &+&
\frac{0.54703}{\Lambda^4} \left( (p_1\cdot p_2)^2+(p_1\cdot
p_3)^2+(p_2\cdot p_3)^2 \right)
\nonumber \\ &+&
\frac{0.70561}{\Lambda^4} \left( p_1^2 \, p_2^2 +p_1^2 \, p_3^2+ p_2^2
\, p_3^2\right)
+\frac{0.40287}{\Lambda^4}\left(  p_1^4+p_2^4+p_3^4\right)
\nonumber \\ &+&
\frac{0.88247}{\Lambda^4} \left( p_1\cdot p_2 \, p_1\cdot p_3+
 p_1\cdot p_2 \, p_2\cdot p_3+p_1\cdot p_3 \, p_2\cdot p_3 \right)
\nonumber \\ &+&
\frac{0.86649}{\Lambda^4}\left( p_1^2 \, p_1\cdot p_2 + p_1^2 \, p_1\cdot p_3
+ p_2^2 \, p_1\cdot p_2  +p_2^2 \, p_2\cdot p_3+
     p_3^2 \, p_1\cdot p_3  +p_3^2 \, p_2\cdot p_3\right)
\nonumber \\ &+&
\frac{0.74448}{\Lambda^4}\left( p_1^2 \, p_2\cdot p_3 +p_2^2 \, p_1\cdot p_3
    +p_3^2 \, p_1\cdot p_2 \right). 
\end{eqnarray}
In the both cases, note the total symmetry in the interchange of 
the momenta $p_1,p_2,p_3$.
To elucidate the effect of this symmetry on the momentum dependence
of the $\gamma 3\pi$ form factor, we re-express the scalar products
$p_i\cdot p_j$ through the Mandelstam variables. We use the definitions
of Ref. \cite{AR96}, which is the Euclidean version of the
definitions in Ref. \cite{Miskimen+al94}:
$s =-(p_1+p_2)^2 \equiv m_\pi^2\bar{s}$, 
$t'=-(p_2+p_3)^2 \equiv m_\pi^2\bar{t}'$, 
$u =-(p_1+p_3)^2 \equiv m_\pi^2\bar{u}$, 
while $ t =-p_3^2 \equiv m_\pi^2\bar{t} $
serves as the measure of the virtuality of the third pion which is
off shell in the CEBAF experiment \cite{Miskimen+al94}. 

On the other hand, in all three pertinent experiments
\cite{Antipov+al87,Miskimen+al94,Moinester+al99}, the first 
two pions {\it are} on shell. We can thus specialize to 
$p_1^2=p_2^2=-m_\pi^2$ and obtain more compact expressions 
for the ${\cal O}(p^4)$ amplitudes in terms of Mandelstam 
variables. For massive quarks Eq. (\ref{amplRA1}) then becomes
\begin{eqnarray}\label{RAMamp1}
\widetilde{F}^{3\pi}_\gamma(s,t',u)&=&\left(0.96274 - 
\frac{ 0.21554 \,m_\pi^2}{\Lambda^2} + 
\frac{0.11534 \,m_\pi^4}{\Lambda^4}\right)
\nonumber \\ & &
+\left( \frac{0.47476 }{\Lambda^2}
- \frac{0.17682\,m_\pi^2}{\Lambda^4}\right)(s+t'+u)
        + \frac{0.15789}{\Lambda^4}(s^2+t'^2+u^2)
\nonumber \\ & &
+ \frac{0.25547 }{\Lambda^4}(s t' + t' u+ s u) 
- \frac{0.08341}{\Lambda^4}(s + t' + u) t
\nonumber \\ & &
 - \left(\frac{0.10777}{\Lambda^2}  -
  \frac{0.07967\,m_\pi^2}{\Lambda^4}\right) t +\frac{0.03776}{\Lambda^
4}t^2  + \frac{0.01000\,m_\pi^2}{\Lambda^4}(m_\pi^2 - t) s   \,   .
\end{eqnarray}
Similarly, in the chiral limit of vanishing $m_\pi$, 
where $p_1^2=p_2^2=0$, the amplitude (\ref{amplRAK1}) becomes
\begin{eqnarray}\label{RAKamp1}
\widetilde{F}^{3\pi}_\gamma(s,t',u)_0&=&
      1  + \frac{0.46114 }{\Lambda^2} (s+t'+u)
      + \frac{0.13676}{\Lambda^4}(s^2+t'^2+u^2)
\nonumber \\ & &
+ \frac{0.22062 }{\Lambda^4}(s t' + t' u+ s u)
- \frac{0.06089}{\Lambda^4}(s + t' + u) t
\nonumber \\ & &
 - \frac{0.08752}{\Lambda^2}  t
+\frac{0.03052}{\Lambda^4}t^2
- \frac{0.00811}{\Lambda^4}  t s  \, .
\end{eqnarray} 
In both cases, we isolated in the last term the violation of 
the $s\leftrightarrow t' \leftrightarrow u$ symmetry, which
occurs when the third pion is off shell, $t\neq m_\pi^2$
(or $t\neq 0$ in the chiral case).

We indicated only the $s,t',u$ dependence of the amplitudes,
since $t$ is of course not independent because of the constraint
$s+t'+u = - p_1^2 - p_2^2 - p_3^2 - q^2$, 
where $q = p_1 + p_2 + p_3$ is the photon momentum.
One can take the photon
to be on shell in all three pertinent $\gamma 3\pi$
experiments~\cite{Antipov+al87,Miskimen+al94,Moinester+al99}.
We thus set $q^2=0$ in addition to  $p_{1}^2=p_{2}^2 = - m_\pi^2$, whereby
the above kinematical constraint becomes
\begin{equation}\label{oscond}
s+t'+u  = 2\, m_\pi^2 + t  \, .
\end{equation}
In any case, this constraint (\ref{oscond})
dictates that the ${\mathcal{O}}(p^2)$-terms, since they
appear in the appropriate symmetric combination, contribute 
only to the part independent of $s,t'$ and $u$. 
This contribution is in fact constant (of the order of
$m_\pi^2$) up to $t$, the virtuality of the third pion.
Therefore, the main contribution to the term linear in $s,t'$ and
$u$ (dominating the $s,t',u$-dependence around the soft limit),
comes from ${\mathcal{O}}(p^4)$ and not ${\mathcal{O}}(p^2)$.
The coefficients of the linear and quadratic terms will thus
be comparably small, giving the parabolic shape to the curves 
displaying our form factors, instead of the steep linear appearance 
\cite{AR96} due to spurious, relatively large linear terms 
(suppressed only as $1/\Lambda^2$) which come from 
${\mathcal{O}}(p^2)$ when there is no symmetry under the interchange 
of the pion momenta [so that the constraint (\ref{oscond}) cannot
do its job].

\vspace*{3mm}

\noindent {\bf 4.}
The experiment which provided the only presently existing 
data point \cite{Antipov+al87} and the one planned at CERN
\cite{Moinester+al99}, belong to the Primakoff type, where
also the third pion is on its mass shell, fixing
$t=m_\pi^2\bar{t}=m_\pi^2$. We use $m_\pi = 138.5$ MeV. 
We then get, in terms of the 
Mandelstam variables expressed in terms of the pion mass squared,
\begin{equation}
\widetilde{F}^{3\pi}_\gamma(s,t') = 1.0319 - 0.00065( \bar{s} + \bar{t'}) 
        +0.00022 ( \bar{s}^2  + \bar{s}\bar{t'} + \bar{t'}^2) \, ,
\label{mARonsh}
\end{equation}
while in the chiral limit, where on shell means $\bar{t} = 0$,
\begin{eqnarray}
\widetilde{F}^{3\pi}_\gamma(s,t')_0 = 1+ 0.000190 (\bar{s}^2  
                     + \bar{s}\bar{t'} + \bar{t'}^2)   \, .
\label{0ARonsh}
\end{eqnarray}
For the second variable fixed to $\bar{t'}=-1$, 
Eqs. (\ref{mARonsh}) and (\ref{0ARonsh}) are depicted,
respectively, by solid and short-dashed curves in Fig. 2.

In the CEBAF measurement \cite{Miskimen+al94}, the third pion 
has spacelike virtuality of the order $t\approx - m_\pi^2$, so we 
also give $\widetilde{F}^{3\pi}_\gamma(s,t')$ obtained by 
fixing $\bar{t} = -1$:
\begin{equation}
\widetilde{F}^{3\pi}_\gamma(s,t') = 
                      0.98524  - 0.00015 \bar{s} - 0.00022  \bar{t'} +
             0.00022 ( \bar{s}^2  + \bar{s}\bar{t'} + \bar{t'}^2 ) \, ,
\label{mARoffsh}
\end{equation}
and again with $\bar{t} = -1$, but in the chiral limit,
\begin{equation}
\widetilde{F}^{3\pi}_\gamma(s,t')_0 = 
                             0.97799 + 0.00022 \bar{s} + 0.00019 \bar{t'} 
+ 0.000190 ( \bar{s}^2  + \bar{s}\bar{t'} + \bar{t'}^2 ) \, .
\label{0ARoffsh}
\end{equation}
CEBAF aims \cite{Miskimen+al94} to measure the $s$-dependence 
of the $\gamma 3\pi$ form factor in the interval 
$s \in [4 m_\pi^2 , 16 m_\pi^2]$, with such kinematics that
$\bar{t'}=-1$ is a good choice for fixing the remaining variable
(as we explained in Ref.~\cite{BiKl99PRD}). 
We thus depict Eqs. (\ref{mARoffsh}) and (\ref{0ARoffsh}) for 
${t'}=- m_\pi^2$ by respective solid and short-dashed curves in Fig. 3.

In Figs. 2 and 3 we also compare our results with some other 
theoretical predictions for ${t}= m_\pi^2$ and ${t}= -m_\pi^2$,
respectively.
The dash-dotted lines represent the chiral perturbation theory 
($\chi$PT) form factor \cite{Bijnens90} (with Holstein's \cite{Holstein96}
choice of renormalization -- {\it i.e.}, we take his \cite{Holstein96} 
Eq. (10) for the $\chi$PT prediction).
The dotted curves are the vector meson dominance (VMD) form factors
\cite{Rudaz84} [i.e., Holstein's \cite{Holstein96} Eq. (9) 
for ${t}= \pm m_\pi^2$]. 

All depicted theoretical form factors indicate that the existing 
data point \cite{Antipov+al87} is probably an overestimate. 
In the considered $s$-interval, the prediction of the present model 
is lower than those of VMD and $\chi$PT. 
The current CEBAF measurement \cite{Miskimen+al94}
should be accurate enough to discriminate between 
at least some of these results.

The most instructive comparison of theoretical predictions 
is the one with the form factors calculated from the box 
graph with the ordinary (``free") constituent quarks looping
inside \cite{BiKl99PRD}. 
In both Fig. 2 and Fig. 3, they are given by the 
long-dashed curve, the line of empty squares and the line 
of crosses, for the constant constituent masses $M$ of 330 MeV
($\approx M_{nucleon}/3$), 400 MeV ($=\sqrt{D}$) and 
580 MeV ($\approx \sqrt{2D}$), respectively. Besides the curves, 
one should also compare the expressions (\ref{RAMamp1})-(\ref{RAKamp1})
for the expansions in powers of scalar momenta $p_i \cdot p_j$ in the 
present~case, with their analogy in our previous paper \cite{BiKl99PRD}. 
One can conclude that
for the presently experimentally interesting momenta, the present
model and the simple ``free" constituent quark loop model agree
quite well as long as the mass scale of these models is similar, 
$\sqrt{2D}\sim M$. The present SD model, with 
its dressed propagators and vertices, does have faster changing
$F^{3\pi}_\gamma(s)$ than the simple constituent quark loop model 
(for the approximately same mass scale, i.e.  $\sqrt{2D}\approx M$),
but this is at the presently considered momenta compensated 
by the larger constant term in the latter model. While we 
can conclude that in the case of this particular form factor,
the present SD model does not bring in the present application 
novel features with respect to the simple constituent quark loop 
model as far as the magnitude of the form factor is concerned,
we can say that considering these two models led to a fairly complete
understanding of the quark box graph calculation of the 
anomalous $\gamma 3\pi$ form factor. 
On the other hand, because of that understanding, the experiment 
can bring an important new input to the SD modeling. If the 
experimental form factor is measured at CEBAF with sufficient 
precision to judge the present SD model results too low, it will
be an unambiguous signal that the SD modeling should be reformulated
and refitted so that it is governed by a smaller mass scale.

\section*{Acknowledgements}
The authors thank R. Alkofer and D. Kekez for many helpful
discussions, K. Kumeri\v{c}ki for checking the manuscript, and 
Physics Department of Bielefeld University for hospitality 
during the work on a substantial part of this paper. 
This work was supported by the Croatian Ministry of Science and Technology 
contract 1-19-222, and B. B. in part by the U.S. Department of Energy 
under Contract No. DE-AC05-84ER40150 and Grant No. DE-FG05-94ER40832.

\newpage

\section*{Figure captions}

\begin{itemize}

\item[{\bf Fig.~1:}] One of the six box diagrams for the process
$\gamma \to \pi^+ \pi^0 \pi^-$. 
Within the scheme of generalized impulse approximation,
                the propagators and vertices are dressed.
The position of the $u$ and $d$
quark flavors on the internal lines, as well as $Q_u$ or $Q_d$
quark charges in the quark-photon vertex, varies from graph to
graph, depending on the position of the quark-pion vertices.

\item[{\bf Fig.~2:}]
Various predictions for the $s$-dependence of the normalized
$\gamma 3\pi$ form factor. We compare
the form factor 
obtained by us with AR Ans\" atze \cite{AR96} for both 
$m_\pi = 138.5$ MeV (solid curve) and the chiral limit
($m_\pi = 0$, dashed curve), with the predictions of the 
vector meson dominance \cite{Rudaz84} (dotted curve),
chiral perturbation theory \cite{Bijnens90,Holstein96} 
(dash-dotted curve), and quark loop model \cite{BiKl99PRD}
for $M=330 \, \mathrm{MeV}$  (long-dashed curve), 
$M=400 \, \mathrm{MeV}$ (boxes), and $M=580 \, \mathrm{MeV}$ (crosses),
and with the experimental data point \cite{Antipov+al87}, 
for all the pions on-shell and $t'=-m_\pi^2$.

\item[{\bf Fig.~3:}]
The comparison of the normalized $\gamma 3\pi$ form factor                    
obtained by us with AR Ans\" atze \cite{AR96} for both                 
$m_\pi = 138.5$ MeV (solid curve) and the chiral limit 
($m_\pi = 0$, dashed curve), with the predictions of the 
vector meson dominance \cite{Rudaz84} (dotted curve),
chiral perturbation theory \cite{Bijnens90,Holstein96} 
(dash-dotted curve), and quark loop model \cite{BiKl99PRD}
for $M=330 \, \mathrm{MeV}$  (long-dashed curve), 
$M=400 \, \mathrm{MeV}$ (boxes), and $M=580 \, \mathrm{MeV}$ (crosses),
for two of the pions on-shell and the third off-shell
so that $t=-m_\pi^2$.
(The Serpukhov data point \cite{Antipov+al87} is also shown 
although it corresponds to all three pions on-shell.)
The remaining variable is again fixed to $t'=-m_\pi^2$.

\end{itemize}


\newpage

\vspace*{4cm}
\epsfig{file=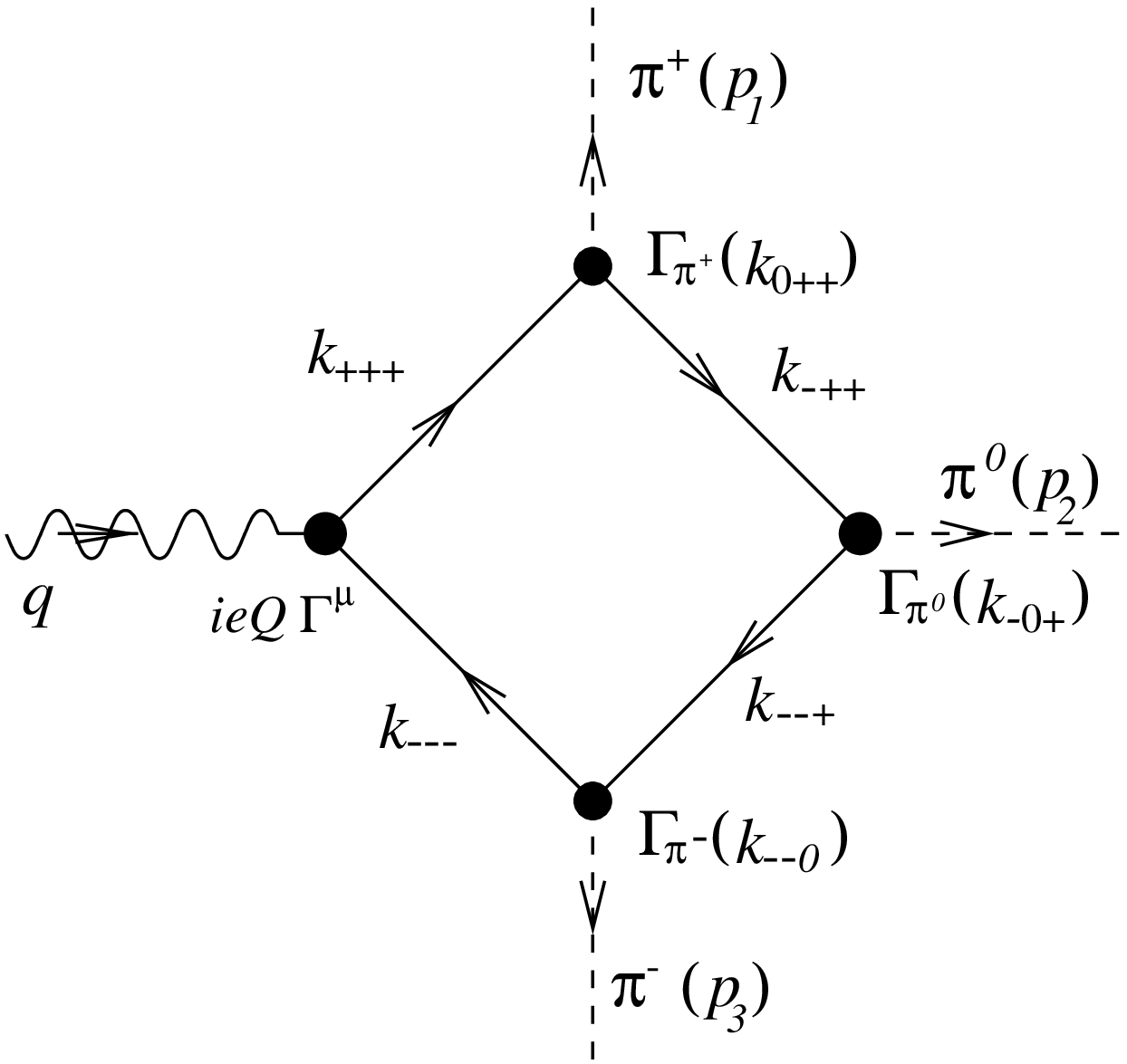}

\newpage

\vspace*{1cm}

\epsfig{file=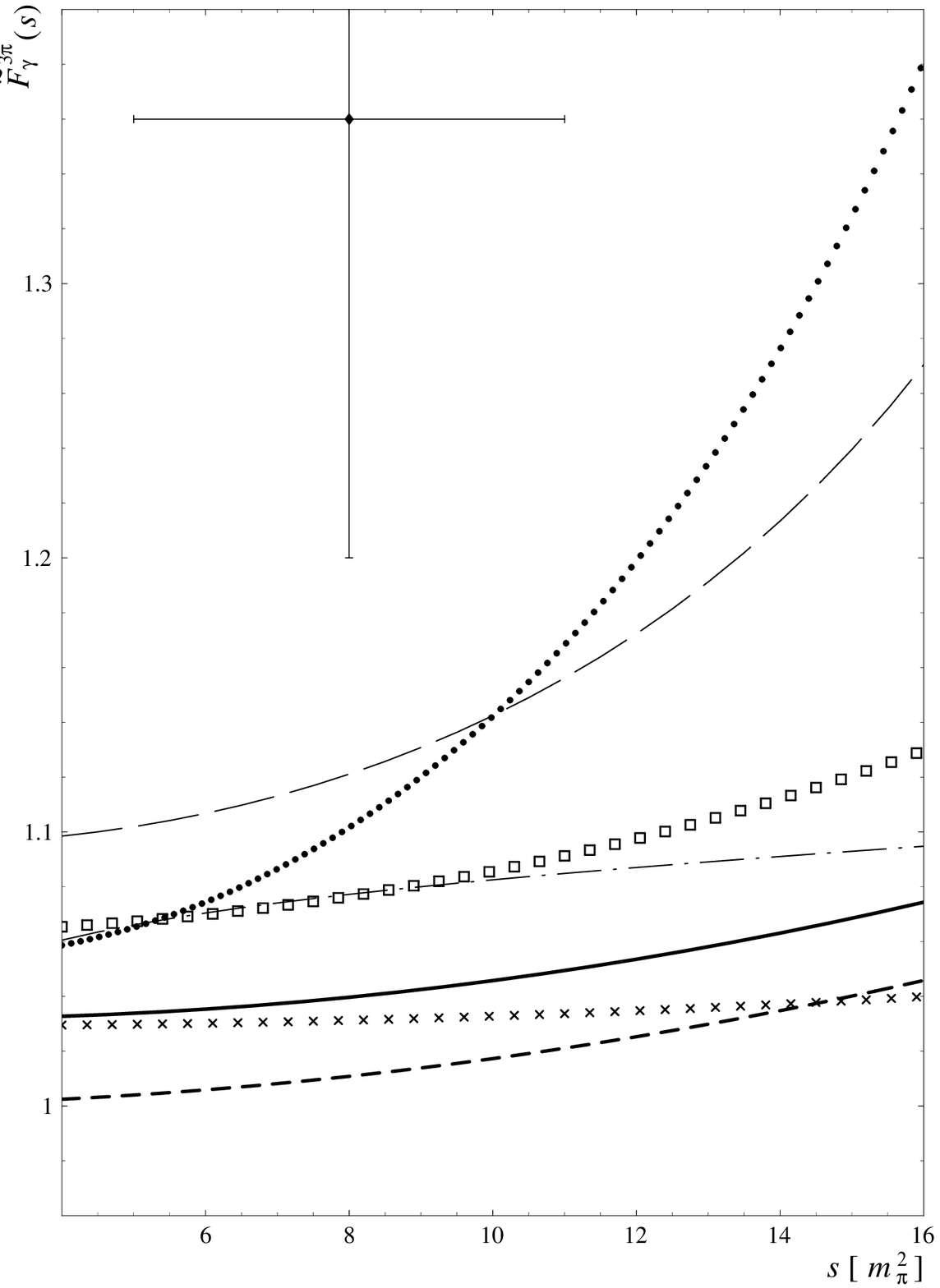,height=21cm}

\newpage
\vspace*{1cm}

\epsfig{file=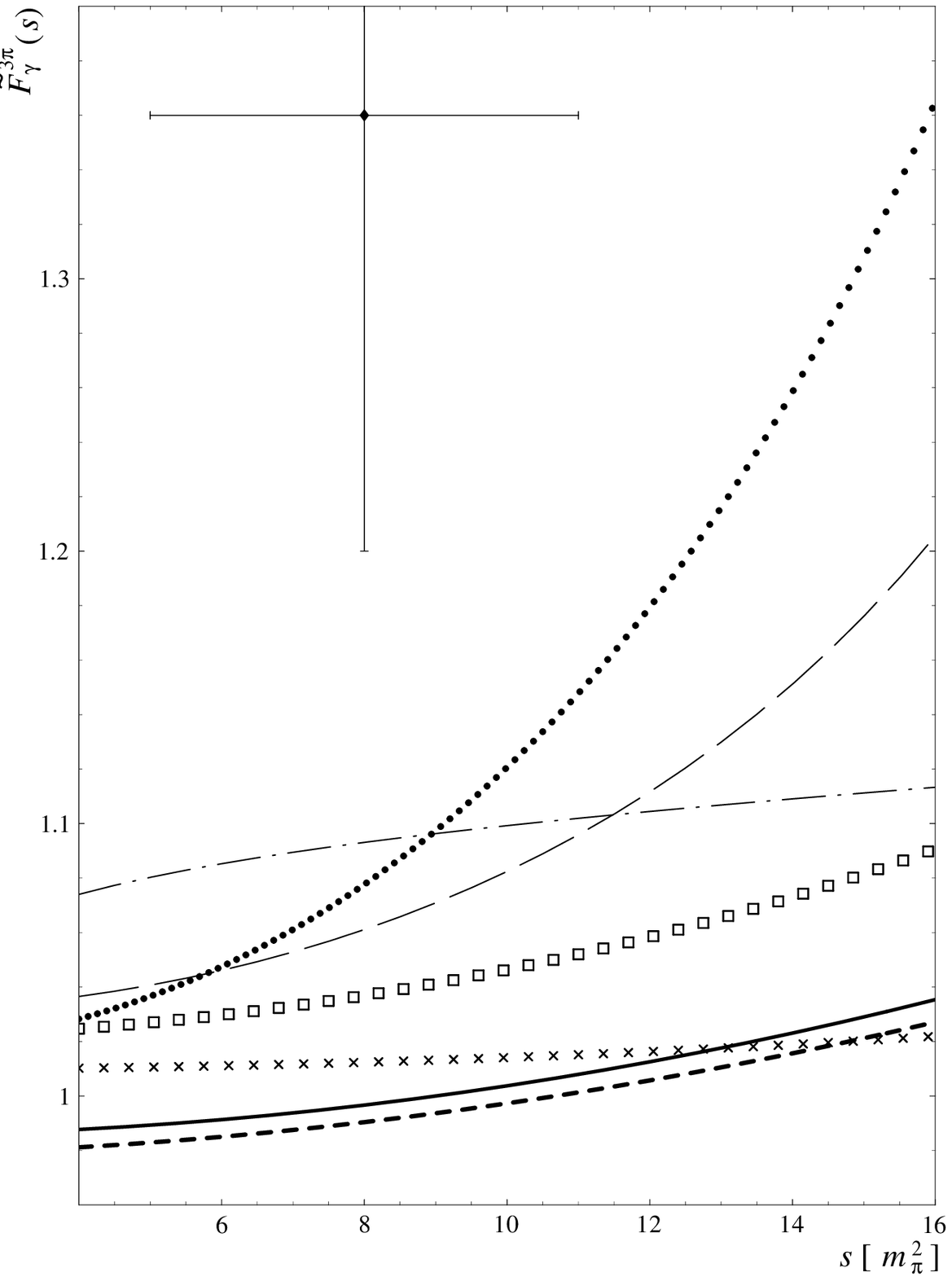,height=21cm}

\end{document}